\documentclass[a4paper,11pt]{article}
\usepackage[legalpaper,margin=1in]{geometry}
\usepackage[hidelinks]{hyperref}
\usepackage{amsmath,amsfonts,amssymb}
\usepackage{xcolor}
\usepackage{graphicx}
\usepackage{float}
\usepackage{authblk}
\usepackage{enumitem}

\DeclareMathOperator{\Tr}{Tr}

\begin{document}

\title{From Quantum Codes to Gravity: A Journey of Gravitizing Quantum Mechanics}

\author[1]{ChunJun Cao}
\affil[1]{Joint Center for Quantum Information and Computer Science, University of Maryland, College Park, MD, 20742, USA.}

\date{}
\maketitle
\begin{abstract}
In this note, I review a recent approach to quantum gravity that ``gravitizes'' quantum mechanics by emerging geometry and gravity from complex quantum states.  Drawing further insights from tensor network toy models in AdS/CFT, I propose that approximate quantum error correction codes, when re-adapted into the aforementioned framework, also has promise in emerging gravity in near-flat geometries.
\end{abstract}
\section{Introduction}

Many familiar approaches to understanding the quantum world begin with theories of classical objects. These classical theories are then quantized to produce their quantum counterparts. Physicists are no stranger to the textbook approaches that take us from a mass on a spring to a quantum harmonic oscillator, or from unassuming spring mattresses to mind-boggling quantum field theories. Given its impressive record of success, it is natural that we also apply the same line of thinking to quantum gravity (QG). 
This has undoubtedly brought about remarkable insight into the heart of QG \cite{dewitt1,dewitt2,polchinski_1998,Rovelli:1997yv}, but we have also encountered daunting challenges and picked up various technical baggage along the way. 
However, if we believe that nature is fundamentally quantum, then applying our classical-centric mindset to study an entirely different quantum ``ecosystem'' would seem like a rather roundabout way to understand the culture of a quantum universe.

In this note, we will try to let go of the classical ideals we know and love and embark on a ``quantum-centric'' journey to QG. Instead of beginning with a classical theory and quantizing it, we take quantum mechanics (QM) as the fundamental theory and emerge semiclassical spacetime descriptions from it. By quantum mechanics, we are referring to the bare-bones quantum theory described by a Hamiltonian $\hat{H}$ and an abstract state vector living in a finite dimensional Hilbert space. While the state contains kinematic information, the Hamiltonian generates dynamics and traverses the Hilbert space in some trajectory\footnote{Equivalently, one can describe a quantum theory as a von Neumann algebra $M$ whereby a state $\rho$ describes the status of the system. }. See \cite{maddog,Cao2017,BEG} for a summary of this approach, which we refer to as the ``gravitize QM'' proposal, and \cite{Giddings2008,giddings2019} for related ideas. 

Let us divide the full gravitize QM problem into 2 parts. (1) From an abstract quantum state, emerge (1a) the semi-classical notion of quantum matter fields on a classical background geometry and (1b) the spatial projection of the Einstein's equations (the Hamiltonian constraint). (2) With additional information from the Hamiltonian $\hat{H}$, emerge the spacetime metric, dynamics, Lorentz invariance, and the full Einstein's equation.

Here we focus on the kinematic problem and provide a partial roadmap for how one can achieve part (1) when given an abstract state vector $|\psi\rangle$. In Section \ref{sec:geometrize}, we first review the approaches that geometrize a quantum state, addressing problem (1a). We then propose that quantum error correction codes can be used as explicit ``emergence maps\cite{emergencemap}'' that allow us to separate the emergent matter field degrees of freedom from the emergent background spacetime. In Section~\ref{sec:gravitize}, addressing problem (1b), we examine an additional constraint on the quantum error correction code that is necessary to emerge linearized Einstein gravity. We will then explain how approximate quantum error correction codes can help us construct such systems. Finally, we briefly comment on open problems in (2) and provide possible directions to pursue in Section \ref{sec:discussion}.


\section{Geometrizing Quantum Mechanics}
\label{sec:geometrize}

The first core problem we tackle is how one can emerge the familiar semi-classical notions of  effective field theory (EFT) on a classical background knowing only a state in the Hilbert space. 

\begin{figure}
    \centering
    \includegraphics[width=0.5\linewidth]{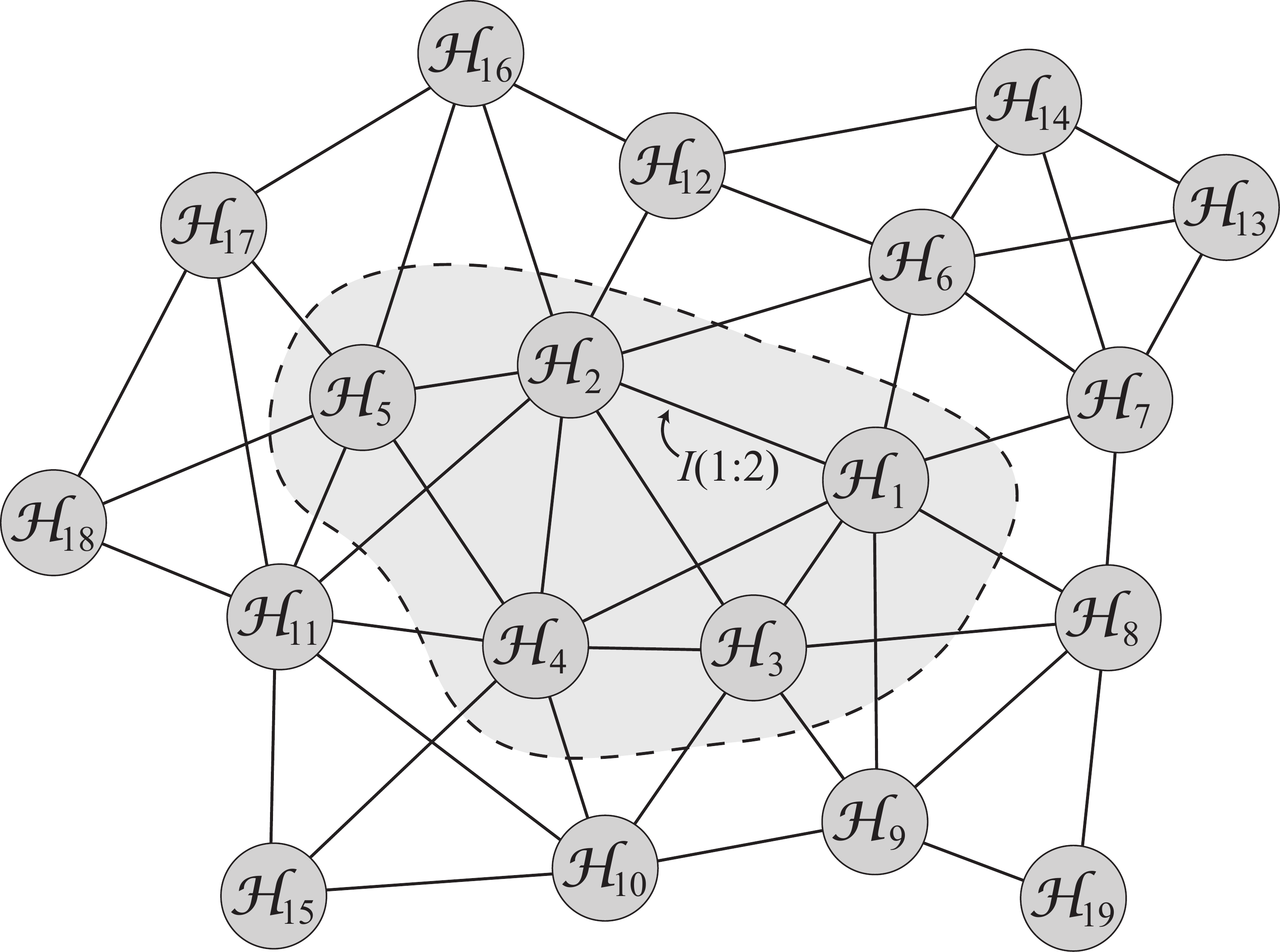}
    \caption{A weighted ``information graph'' created from a quantum state living in a factorizable Hilbert space.}
    \label{fig:rcgraph}
\end{figure}

\begin{figure}
    \centering
    \includegraphics[width=0.7\textwidth]{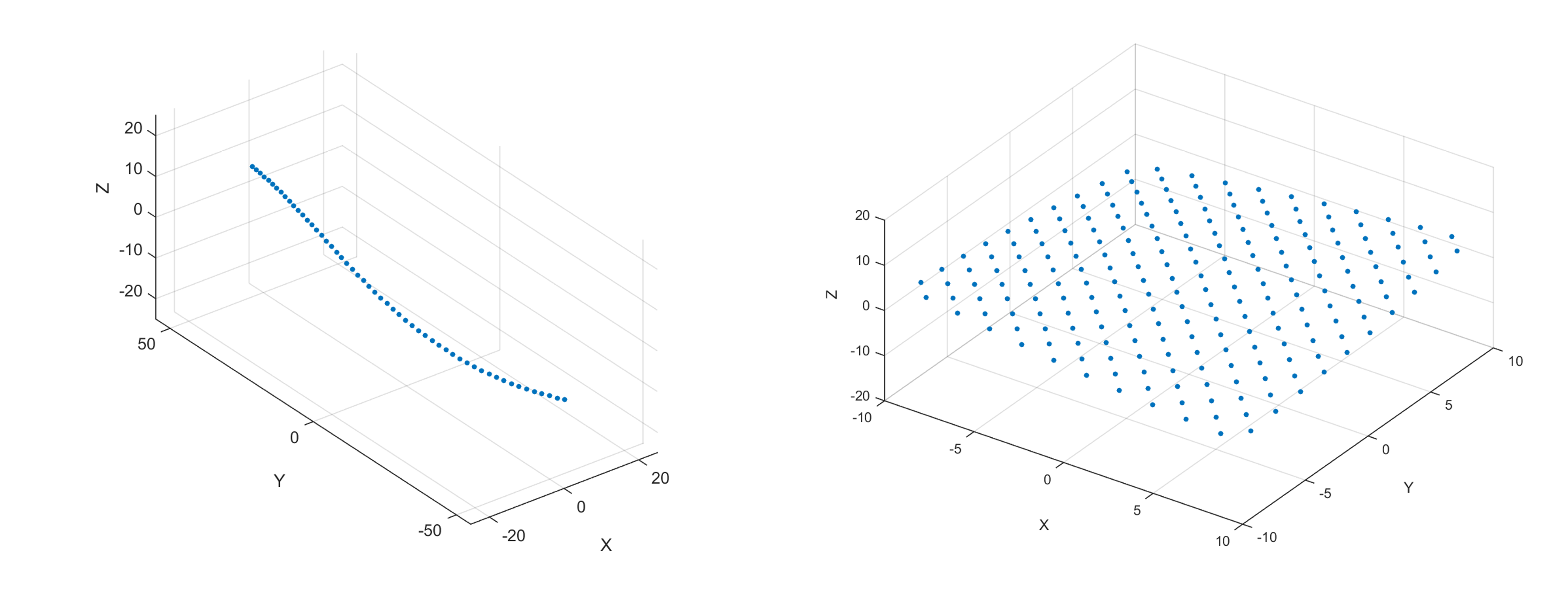}
    \caption{For each quantum state, one can try to determine the emergent geometry using multi-dimensional scaling, which rearranges the vertices of the ``information graph'' according to their relative distances. Two resulting reconstructions from the algorithm are shown, where each blue dot is a graph vertex. }
    \label{fig:mds}
\end{figure}

Recall from \cite{Cao2017} that by having a sufficiently ``nice'' state\footnote{Also known as a redundancy constrained state, which is roughly a graph generalization of an area-law state.} in a Hilbert space with known factorization $\mathcal{H}=\bigotimes_i \mathcal{H}_i$, 
one can convert the state $|\psi\rangle\in\mathcal{H}$ into a graph where the vertices of the graph (Figure~\ref{fig:rcgraph}) correspond to the tensor factors and their edges are weighted by the mutual information 
\begin{equation}
    I(i:j) = S(i) + S(j)-S(i\cup j).
\end{equation}
Here $S(i)$ is the von Neumann entropy of the reduced state on tensor factor $i$. 
If we assume that the parts with high mutual information are closer together and the parts with low mutual information are farther apart, one can define a notion of distance and recover the best-fit dimensionality of the emergent geometry. We can also rearrange the tensor factors in a geometrical fashion (Figure~\ref{fig:mds}) using multi-dimensional scaling\cite{mdsbook,gmds}. If these vertices can be (near)-isometrically embedded in a Riemannian manifold, then we declare the target manifold as the (approximate) emergent spatial geometry. For emergent geometries that are perturbatively close to flat manifolds, it is also possible to recover the full spatial metric $g_{ij}$ using inverse tensor Radon transform\cite{BEG,sharafutdinov1994integral,Monard2014,Monard2015}. 

Although this method allows us to reconstruct the (approximate) emergent geometry, it does not tell us how to separate a quantum field on this geometry from the background itself. To do so, we require an emergence map \cite{emergencemap} that can distinguish the matter, or effective field theory (EFT), degrees of freedom from those that build up the background geometry. Here we claim that the encoding map of a quantum error correction code can act as such an emergence map and can also provide practically useful relations for our purposes. Note that this does not modify the bulk entanglement gravity framework\cite{BEG} but simply provides a more concrete implementation. 

\subsection{A Primer for Quantum Error Correction Codes}
Let us first review some basic properties of quantum error correction codes (QECC)\cite{gottesman,lidar_brun_2013} that will be essential for the rest of our discussion. 
Generally, we define a quantum code as a Hilbert subspace
\begin{equation}
\mathcal{C}\subset\mathcal{H},
\end{equation}
where $\mathcal{H}$ is the ``physical'' Hilbert space, which intuitively can be understood as a tensor product of qubits that one manipulates in a laboratory setting. $\mathcal{C}$ is called the code subspace; states in this subspace usually satisfy certain desirable properties such that the quantum information (or logical information) they encode is protected against errors that may occur on the physical qubits. On a more abstract level, the physical qubits, which are tensor factors of dimension 2, can be replaced by the more general tensor factors $\mathcal{H}_i$ in the previous section.

More conveniently, one can keep track of the code subspace by defining a quantum code as a linear map $V:\mathcal{H}_{\rm logical}\rightarrow \mathcal{H}$  that smears the logical information in $\mathcal{H}_{\rm logical}$ over a larger physical Hilbert space, where $\mathcal{H}_{\rm logical}$ is isomorphic to $\mathcal{C}$. If we assume the encoding process to be unitary then $V$ is also an isometry. We refer to $V$ as the encoding map of a code.

\begin{figure}
    \centering
    \includegraphics[width=0.4\linewidth]{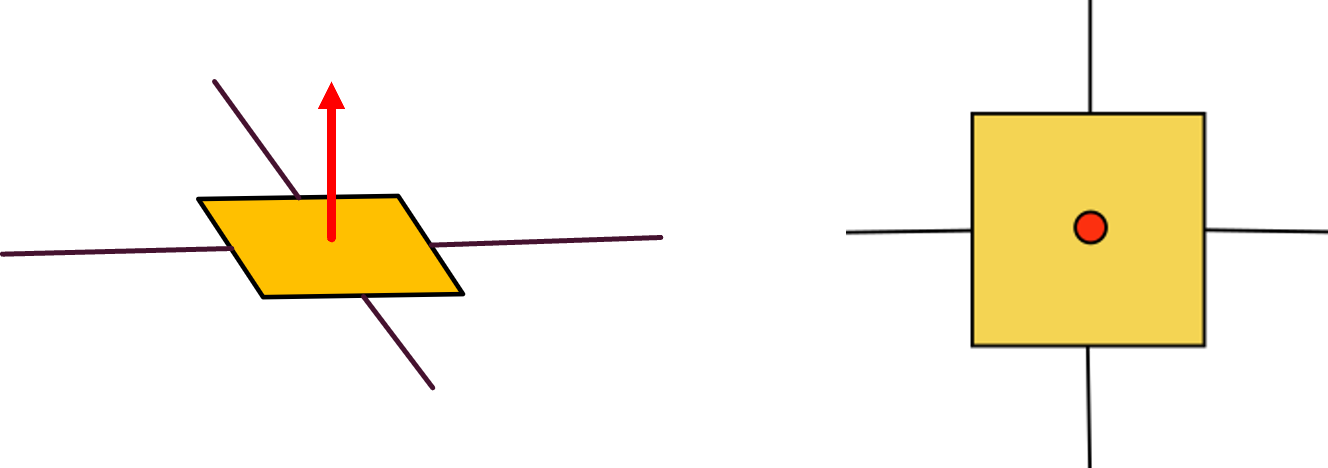}
    \caption{Tensor network representation of the 4 qubit code. The red dangling leg represents the single qubit input degree of freedom while the 4 in-plane legs represent the 4 physical qubits in the output Hilbert space. An equivalent figure on the right uses a red dot to denote the logical qubit when the tensor when viewed from top down.}
    \label{fig:4qubitcode}
\end{figure}

Similar to classical error correction codes where 0 and 1 are encoded as bit strings over multiple bits, the quantum code has encoded quantum states as entangled states over many qubits. For a simple example, consider a 4-qubit code (Figure~\ref{fig:4qubitcode}) defined by the following encoding isometry
\begin{equation}
    V \propto |0000\rangle\langle 0|+|1111\rangle\langle 0| +|1100\rangle\langle 1| +|0011\rangle\langle 1|.
\end{equation}
This map smears the logical information of a single qubit onto 4 qubits entangled in a GHZ state, where we may encode a single qubit state $|\psi\rangle = a|0\rangle+b|1\rangle$ as an encoded states $|\tilde{\psi}\rangle$ over 4 qubits. 
\begin{equation}
|\tilde{\psi}\rangle = a|\tilde{0}\rangle +b|\tilde{1}\rangle,~ \textrm{where}~
    |\tilde{0}\rangle \propto |0000\rangle+|1111\rangle, ~|\tilde{1}\rangle \propto |1100\rangle+|0011\rangle. 
\end{equation}
 We will distinguish all encoded states with tilde throughout the note. A tensor network representation of this simple code is shown in Figure~\ref{fig:4qubitcode}.

Although it is easy to define a quantum code, it is much harder to find a good quantum code that protects encoded information against errors. A quantum code is also an error correction code when the effects of certain errors can be undone. 

In the 4-qubit code example, consider an erasure error where one qubit is ``lost''. Assume, without loss of generality, that the 4th qubit is lost, and we can only perform operations on qubits 1 through 3 to recover the encoded information. Here it is easy to show that there is an explicit unitary $U_{123}$ independent of the state $|\tilde{\psi}\rangle$ such that

\begin{equation}
    U_{123}\otimes I_4|\tilde{\psi}\rangle = |\psi\rangle_1 |0\rangle_2 |\chi\rangle_{34}
    \label{eqn:4qbrec}
\end{equation}
where $|\chi\rangle_{34}$ is a Bell state\cite{Cao:2020uvb}. In this case, the erasure error of a single qubit is correctable because the encoded information $|\psi\rangle$ is undamaged by the erasure. Therefore, the 4-qubit code is a QECC that corrects single qubit erasure errors.

The existence of such a decoding unitary or, more generally, a recovery map is a defining feature of QECCs at large. \cite{Harlow:2016vwg} showed that for any code where the physical Hilbert space can be factorized as $\mathcal{H}=\mathcal{H}_B\otimes \mathcal{H}_{\bar{B}}$ and that the erasure of subsystem $\bar{B}$ is correctable, then 
there exists decoding unitary $U_B$ such that 
\begin{equation}
    U_B\tilde{\rho}_B U_B^{\dagger} = \rho_{B_1}\otimes \chi_{B_2}
\end{equation}
where $\mathcal{H}_{B_1}\otimes \mathcal{H}_{B_2} \cong \mathcal{H}_B$, and $\tilde{\rho}_B=\Tr_{\bar{B}}[|\tilde{\psi}\rangle\langle\tilde{\psi}|]$. 

Assuming the existence of recovery procedures provided by such unitaries $U_B$, which is expected for generic QECCs\footnote{By generic, we mean that something like this is typically true up to small corrections for a QECC whose code subspace is small compared to the physical Hilbert space and that the encoding unitary is drawn from the Haar measure\cite{HaydenPreskill}. }, we can naturally split an encoded state $\tilde{\rho}$ into two parts --- one containing the encoded information that we wish to recover and the other capturing the underlying entanglement necessary a code to have non-trivial error correction properties. One can extract an entropic relation from these pieces of information by computing the von Neumann entropies of the quantities on both sides of the equation. Note that the entropy is invariant under unitary conjugation, therefore 
\begin{equation}
    S(\tilde{\rho}_B) = S(\rho_{B_1})+S(\chi_{B_2}),
    \label{eqn:qeccrt}
\end{equation}

where $S(\rho_{B_1})$ corresponds to the entropy of the encoded information that is recoverable on $B$. This can be nonzero even when one encodes a pure state because generally a QECC can encode multiple logical degrees of freedom from which operations on $B$ can only recover a fraction of them. Therefore, when the encoded state is entangled, $S(\rho_{B_1})$ represents the entanglement entropy of the reduced state recovered from $B$. $S(\chi_{B_2})$ is the leftover entanglement that is essential in building a non-trivial quantum error correction code. In our 4-qubit example (\ref{eqn:4qbrec}), if we take $B$ to be the union of qubits 1 through 3 and $\bar{B}$ to be the erased 4th qubit, then  $\rho_{B_1}=|\psi\rangle\langle \psi|$ is the information we extracted on qubit 1 after decoding (\ref{eqn:4qbrec}). We obtain $\chi_{B_2}$ by tracing out the 4th qubit from the state $|0\rangle_2|\chi\rangle_{34}$, such that  $\chi_{B_2}=|0\rangle\langle 0|\otimes I/2$ is supported on qubits 2 and 3. In this case, $S(\rho_{B_1})=0, S(\chi_{B_2})=1$. For typical QECCs, when $\log \dim\mathcal{H}\gg \log\dim\mathcal{C}$, $S(\chi_{B_1})\gg S(\rho_{B_1})$.

Note that expressions like the above entropy formula holds more generally as long as such ``decoding unitary'' exists. For instance, they hold for certain approximate erasure correction codes\cite{ABSC} in a state-dependent fashion. More recently, this is made precise by \cite{Akers:2021fut} in a more comprehensive statement, which further extended the theorem by \cite{Harlow:2016vwg} to a wider class of quantum codes\footnote{There may be added benefits in having the code to have complementary recovery properties (defined in \cite{Harlow:2016vwg}), but it is unclear at this point whether that is necessary for our goals in this note. }.

\subsection{What does it all mean for quantum gravity?}
So how does QECC function as an emergence map which separates matter from the background geometry? For pedagogical reasons, let us first acquire some intuitions from AdS/CFT where concrete connections have been made. We will find that (\ref{eqn:qeccrt}) is analogous to the familiar notion of generalized entropy, where $S(\rho_{B_1})$ is the matter entropy contribution while $S(\chi_{B_2})$ comes from the area of some surface.



Recently, \cite{Almheiri:2014lwa} showed that AdS/CFT has a strong resemblance with (approximate) quantum erasure correction codes, where the code subspace $\mathcal{C}$ is defined as the bulk low energy subspace of the theory. The physical processes in this low energy regime can be approximated by an effective field theory that lives on the AdS background in the bulk. See \cite{Harlow:2018fse} for a detailed review of the subject. 

The physical Hilbert space $\mathcal{H}$ in this context corresponds to the boundary (CFT) degrees of freedom. The code subspace is spanned by a comparatively small number of low energy states of the EFT that do not cause significant gravitational backreactions. If we limit ourselves to the regime where the approximation of quantum field theory on curved spacetime is valid, then the encoded ``logical qubits'' in the code subspace correspond to the matter field degrees of freedom floating on top of a fixed background geometry while choosing a code subspace amounts to picking the said common background geometry on which the matter field lives.

\begin{figure}
    \centering
    \includegraphics[width=0.95\linewidth]{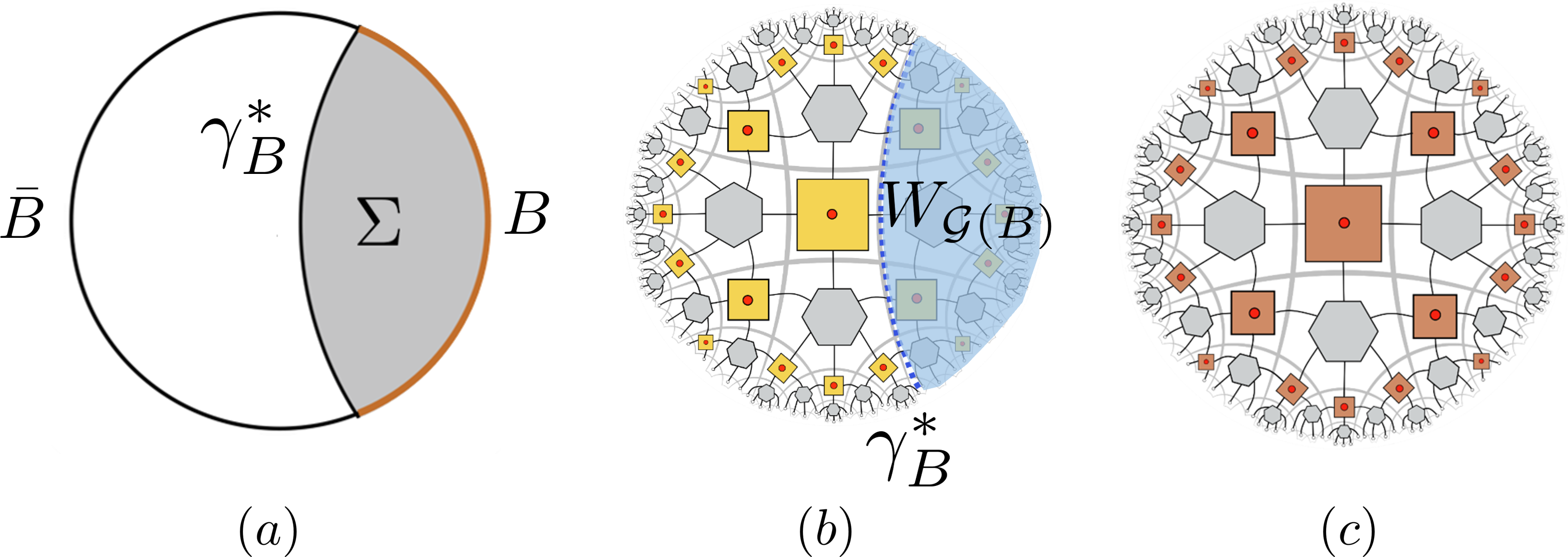}
    \caption{(a) A time slice of AdS. Shaded region represents an entanglement wedge of $B$. The bulk or encoded degrees of freedom in the entanglement wedge of $B$ can be recovered from $B$. (b) An analogous picture of (a) in terms of a holographic tensor network, which is a graphical representation of an encoding map. The shaded entanglement wedge of $B$ can be similarly defined through decoding operations. A ``geodesic'' $\gamma_B^*$ is defined as the boundary of that wedge. (c) A ``noisy'' version of the holographic code to the left, where each of the 4-qubit codes that made up of the tensor network are replaced by the skewed, approximate QECC counterparts.}
    \label{fig:holography}
\end{figure}

If we select some physical subsystem in $\mathcal{H}_B$ (Figure~\ref{fig:holography}a) that corresponds to a boundary subregion $B$, one can extract the encoded information from the reduced state on $B$. The entropy of the recoverable logical information $S(\rho_{B_1})$ now takes on a specific meaning --- it is the entropy of the matter (fields) in a subregion of the bulk AdS known as the entanglement wedge. More precisely, we are referring to the vacuum-subtracted entropy of the matter field in a subregion \cite{Casini:2008cr,Bousso:2014sda,Bousso:2014uxa}. The entanglement wedge of $B$ is shown in Figure~\ref{fig:holography}a as the shaded region, which is bounded by the boundary segment of $B$ and a bulk spacelike geodesic $\gamma_B^*$ that is anchored on the endpoints of $B$. As a QECC, the bulk/logical degrees of freedom in the EFT in $\Sigma$ is recoverable via operations that only have support on $B$.

As we mentioned, choosing the code subspace effectively selects the background geometry. Once we have extracted the encoded information $\rho_{B_1}$ from a recovery map with support on $B$, the remaining information that goes into defining the code, and hence the geometry, is captured by the state $\chi_{B_2}$. The entropy $S(\chi_{B_2})$ generally holds the most of the entanglement between subsystems $B$ and $\bar{B}$ when $\log\dim\mathcal{C}\ll \log\dim\mathcal{H}$. Intuitively, this entropy contribution is the part that is responsible for emerging the space-time geometry. In the tensor networks, it is roughly given by the number of graph edges the geodesic $\gamma_B^*$ cuts through in the bulk (Figure~\ref{fig:holography}b). This defines a notion of area of a minimal surface (or length of a geodesic in the 2 dimensional picture) that separates one spatial region subtended by $B$ from an adjacent region subtended by $\bar{B}$.


Indeed, in the ``continuum limit'' where we have the actual AdS/CFT correspondence instead of a discrete toy model, this second entropy term becomes the area of the minimal surface where the precise the holographic counterpart\cite{Faulkner:2013ana} of (\ref{eqn:qeccrt}) reads 
\begin{equation}
    S(B) = S_{\rm matter}+S_{\rm geom} = S(\Sigma)+ \frac{\mathcal{A}_B}{4G_N}
\end{equation}
In the regime where gravity is weakly coupled, the dominant contribution to $S(B)$ is $S_{\rm geom}\propto\mathcal{A}$, the area of the minimal surface separating B from $\bar{B}$ (or $\Sigma$ from its complement in the bulk), which contains the geometric information of the background. This is again analogous to what we found in the generic QECC story above when the code subspace is much smaller than the physical Hilbert space. In Figure~\ref{fig:holography}a, $\mathcal{A}$ is simply the length of $\gamma_B^*$. The matter entropy $S_{\rm matter}$, coming from the bulk fields in the region $\Sigma$ is a subleading contribution in the large $N$ expansion.



\subsection{Roadmap for General Code Geometrization}
If we think of AdS/CFT as a special implementation of a QECC that happens to have an emergent near-hyperbolic geometry, then we can also imagine constructing a QECC with other emergent geometries that are not AdS, where it would be similarly reasonable to identify $\rho_{B_1}$ with the state of the matter field in a spatial subregion and $S(\chi_{B_2})$ with the interface area of two adjacent regions in that emergent geometry.

More specifically, let us conjecture that for any quantum code where the entropy formula (\ref{eqn:qeccrt}) can be defined via a recovery map induced by the decoding unitary,
the generalized entropy is given by the von Neumann entropy of a subsystem $B$ in the physical Hilbert space. The matter entropy is given by that of the encoded information and the entropy which goes into defining  the ``background geometry'' is the entropy required for building up a non-trivial code,
\begin{equation}
    S_{\rm gen}:=S(B), ~ S_{\rm matter}:= S(\rho_{B_1}),~\mathrm{and}~  S_{\rm geom} :=S(\chi_{B_2}).
\end{equation}

Furthermore, the ``geometric'' portion of the entropy $S_{\rm geom}$ can take on the meaning of the area of a spacelike surface\cite{Ryu:2006bv,BEG} that separates two adjacent regions  when the quantum state admits an emergent geometry, i.e.

\begin{equation}
    S_{\rm geom} \propto\frac{\mathcal{A}_B}{4G_N}.
\end{equation}

Let us summarize the minimal assumptions for the partial roadmap we have just built for (1a). For this part, we assume that

\begin{enumerate}

    \item The Hilbert space is factorizable
    \begin{equation}
    \mathcal{H}=\bigotimes_i \mathcal{H}_i.
    \end{equation}

    \item We are given a preferred subspace  
    \begin{equation}
        \mathcal{C}\subset \mathcal{H}
    \end{equation}
    which defines a quantum code. Equivalently, we can be given a linear encoding map $V$ which identifies the code subspace. 
    
    \item For any state $|\tilde{\psi}\rangle \in \mathcal{C}$ and for some choices of subsystem $B$, the quantum code admits a decoding unitary $U_B$  which induces the decomposition of the von Neumann entropy
    
    \begin{equation}
        S(\tilde{\rho}_B) = S(\rho_{B_1})+S(\chi_{B_2})
    \end{equation}
    
        \item The \textit{geometric mutual information}
        \begin{equation}
            I_G(B:\bar{B}) := S(\chi_B)+S(\chi_{\bar{B}})- S(\chi_{B\cup \bar{B}})
        \end{equation}
        of  subsystems $B$ and $\bar{B}$ is proportional to the interface area of two adjacent regions in the emergent geometry, if it exists
    \begin{equation}
        \mathcal{A}(B,\bar{B}) \propto I_G(B:\bar{B}).
    \end{equation}
    
    \item The logical information $|\psi\rangle\in \mathcal{H}_{\rm logical}$ corresponds to the state of the emergent matter field on a fixed background geometry consistent with the area data $\{\mathcal{A}\}$. This implies that $S(\rho_{B_1})$ is contributed by the matter field.
    
\end{enumerate}

Let us briefly examine how these assumptions may be derived or justified.
A finite dimensional Hilbert space is often factorizable\footnote{Strictly speaking, we do not even need to assume a factorization as long as the relevant entropic quantities are known because the techniques in \cite{BEG} only truly make use of the entropies. Nevertheless, we will leave this generalization for future discussion.} as long as does not have prime dimensionality\footnote{However, when it is not factorizable, it is also possible to consider decompositions in which a subspace of it is \cite{Singh2018}.}. Note that the tensor factorization may not be unique, and often there is a preferred factorization for a quantum system. Different approaches may be deployed in finding this decomposition, which involves detailed information of $\hat{H}$ or von Neumann algebra $M$ such that the emergent dynamics is ``quasi-classical''\cite{mereology}. In some cases, there can be multiple valid factorizations that satisfy this requirement. Although such systems exist, e.g. systems with dualities \cite{coleman, Maldacena:1997re}, they are exceedingly rare\cite{Cotler:2017abq}.


Building a non-trivial quantum code that admits reasonable decoding from different subsystems $B$ and $\bar{B}$ is of both theoretical and practical importance.
Nevertheless, the required property (c.f. Assumption 3) is still sufficiently generic and can be found in many popular constructions like (approximate) stabilizer codes. In fact, from the analysis of random codes\cite{Hayden:2016cfa}, which can be understood as characterizing the typical behaviours of quantum codes, there is reason to believe that it should hold for most QECCs whose code subspace is much smaller than the physical Hilbert space.

The identification of the matter and geometric entropies are motivated from AdS/CFT. For pure states, which we focus on in this note, it is equivalent to define the interface area as $S(\chi_B)$ or mutual information as the two quantities are identical up to a factor of 2. There is also reason to expect that codes roughly satisfying such assumptions or emergent geometries exist. For instance, one can easily construct such (approximate) QECCs using random tensor networks \cite{Hayden:2016cfa}. An exact construction for a code with flat geometry is also possible \cite{Cao:2021ibt}. 

Using assumptions (2), (3) and (5), we can now easily isolate the EFT degrees of freedom from the background via decoding and recovering the logical state. Because $|\psi\rangle$ as a logical state over a number of logical qubits corresponds to the quantum state of the EFT, ideally it should reflect the low energy state entanglement patterns of a quantum field theory on a particular background.

To obtain the background geometry itself, we note that much of these techniques, which we reviewed earlier on, have already been developed for near-flat geometries. The only difference is that the generalized entropy $S(\tilde{\rho}_B)$ was used to obtain an approximate emergent geometry. With the information given by assumptions (1), (3) and (4), we can perform a more careful reconstruction using the actual geometric contribution $S(\chi_{B_2})$ given by the QECC emergence map. Then we simply replace $S(\tilde{\rho}_B)$ by $S(\chi_{B_2})$ in all of the computations in \cite{Cao2017,BEG}. Note that under the typical expectation where $\log\dim\mathcal{C}\ll \log\dim\mathcal{H}$,  this replacement does not significantly alter the resulting geometry as $S(\chi_{B_2})$ is dominant and $S(\rho_{B_1})$ is subleading. However, we can expect the newer reconstruction to be better, especially when the matter field $|\psi\rangle$ contains entanglement that is non-local with respect to the background geometry. By repeating the procedures in \cite{Cao2017}, we can again obtain the full spatial metric tensor $g_{ij}$. Note that the set of entropy data $\{S(\chi_{B_2}), \forall B\}$ need not be always be consistent with a classical Riemannian geometry with metric $g_{ij}$. However, this non-geometricity may be quantified with the range characterization of the tensor Radon transform \cite{sharafutdinov1994integral,Monard2014,Monard2015}, which has recently been implemented in the context of AdS/CFT\cite{Cao:2020uvb}. The problem is even easier for near-flat backgrounds since both the reconstruction formula and the range characterization have been known for quite some time now.

\section{Gravitizing  Quantum Mechanics}
\label{sec:gravitize}
\subsection{Constraints for Linearized Gravity}
The roadmap thus far covers how one can go from a set of abstract, amorphous quantum states to a semi-classical picture where quantum matter is living on top of a background geometry. In practice, the explicit process is most well-understood when the emergent geometry is also near-flat. However, more is needed to emerge gravity as we have not explained how the emergent matter and emergent geometry should interact with each other in the picture so far. 


Suppose we have a QECC with encoded state $|\tilde{\psi}\rangle$ from which we have obtained an emergent flat geometry with quantum matter in the vacuum of some EFT. Now consider perturbing the state $|\tilde{\psi}\rangle \rightarrow |\tilde{\sigma}\rangle\in \mathcal{C}$. Intuitively, if $|\tilde{\psi}\rangle$ is the vacuum state, then we have changed the encoded state to an excited state with a different matter distribution in the emergent geometry. Now from gravitational intuitions, we also expect the matter excitation to back-react and alter the background geometry. Such behaviour is made precised by the Einstein's equations.

Because we are dealing with spatial geometries in this note, let us recast the Einstein's equations in a form that can be related to our current discussion. Consider the linearized Einstein's equations on flat spacetime; we can project it onto a single timeslice with a timelike unit normal $t^{\mu}$, which produces a linearized Hamiltonian constraint (LHC). Note that if the system satisfies the LHC for all $t^{\mu}$ and is Lorentz invariant, then it will also satisfy the full linearized Einstein's equations. Therefore, reproducing these constraints will be a key step towards emerging gravity. It was shown in \cite{BEG} that the LHC against flat background can be translated into an entropic constraint 
\begin{equation}
    \delta \mathcal{R}=16\pi G_N\delta T_{tt} \iff \delta S_{geom}[\delta g_{ij}] +\delta S_{matter}[\delta T_{tt}] =0,
    \label{eqn:hamconstr}
\end{equation}
where $\mathcal{R}$ is the spatial curvature, $T_{tt}$ is the $tt$-component of the stress-energy tensor, and $G_N$ is the gravitational constant. 
This implies that linearized gravity imposes an additional constraint on the QECC, which is expected. This relation also makes intuitive sense because changes in the matter field excitations lead to changes in $\delta T_{tt}$, which then cause changes in $\delta S_{matter}[\delta T_{tt}]$. Through the above entropic relation, they incur perturbations in $\delta S_{geom}$ which can be related to changes in the area of the interface $\delta \mathcal{A}$ through our Assumption 4. $\delta \mathcal{A}$ can then be converted to changes in the background metric $g_{ij}$ by solving an inverse problem in the tensor Radon transform.

\subsection{The Need for Approximate QECC}





However, (\ref{eqn:hamconstr}) is not satisfied by any known stabilizer QECC models. For example, in the stabilizer tensor network toy models for holography \cite{Pastawski:2015qua, ABSC}, it is easy to check that $\delta S_{\rm geom}=0$ for any $\delta S_{\rm matter}$. At a first glance, this is expected of any code that can correct erasure errors perfectly because if $S(\chi_{B_2})$ depends on the logical information, then one can learn something about the encoded information on $B$ by measurements\footnote{Here we are simply stating what appears to be a reasonable observation, rather than a theorem. In fact, if the code subalgebra supported on $B$ has a non-trivial center, then both $B$ and $\bar{B}$ can have some degree of access to the same classical information. In that case, the erasure of $\bar{B}$ does not damage the encoded information on $B$ because classical information can be cloned.}, e.g. Renyi entropies, from $\bar{B}$. It would then appear that the encoded information is not robust against erasure errors on $\bar{B}$.  As such, existing holographic QECC models, such as the one shown in Figure~\ref{fig:holography}, best correspond to the picture of quantum field theory on a fixed background, where matter is decoupled from gravity.

Therefore, to find a code with the above property of ``gravitational backreactions'', we now look to quantum codes whose encoded information is less robust against errors. These approximate QECCs (AQECC) are usually less desirable for the purpose of building fault-tolerant quantum computers, but are incredibly useful here. This is unsurprising for experts in AdS/CFT because the holographic QECC has long been expected to be an approximate erasure error correction code once gravitational effects are included\cite{Almheiri:2014lwa}.

We can easily construct some of these approximate QECCs --- they are still very much quantum codes by our earlier definition, except some errors can now only be corrected approximately. Consider a ``skewed'' version of the 4-qubit code we saw earlier,
\begin{equation}
    |\tilde{0}\rangle = \frac {1}{\sqrt{2}}(|0000\rangle +|1111\rangle), ~
    |\tilde{1}\rangle = \alpha |1100\rangle +\beta|0011\rangle. 
\end{equation}

It corresponds to the 4-qubit code when $\alpha=\beta$, which corrects any single qubit erasure exactly. However, that erasure correction is only approximate if $\alpha \approx \beta$. We can end up with such a code if the quantum computer suffered from correlated coherent noise that acts as a global unitary on all 4 qubits during the encoding process. 
\begin{figure}
    \centering
    \includegraphics[width=0.95\linewidth]{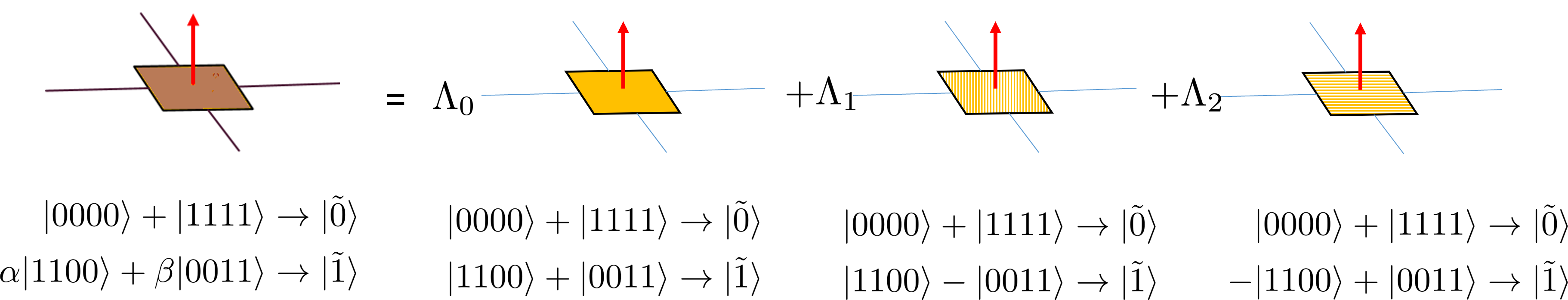}
    \caption{Tensor of a noisy 4-qubit code and how it breaks down as a superposition of different stabilizer codes. The corresponding encoding maps are given below each figure.}
    \label{fig:4skewed}
\end{figure}

More generally, we can express these approximate or skewed codes as a linear encoding map $V_{\epsilon}:\mathcal{H}_{\rm logical}\rightarrow \mathcal{H}$
where we perturb a good QECC, defined by the encoding map $V_0$, with some ``noise'' terms $V_{i\ne 0}$ such that
\begin{equation}
    V_{\epsilon}=\Lambda_0 V_0+\sum_{i\ne 0}\Lambda_i V_i, ~~\sum_{i}|\Lambda_i|<\epsilon,
    \label{eqn:nonisoencode}
\end{equation}

where each $V_i$ is an encoding map of a QECC different from $V_0$. For instance, the 4-qubit approximate code above can be broken down as a superposition of encoding isometries shown in Figure~\ref{fig:4skewed} with some choice of $\Lambda_i(\alpha)$. This decomposition into a superposition of encoding maps is highly non-unique. However, when the perturbations are small, it is most natural to identify the ``noiseless'' QECC encoding map $V_0$ that has the largest overlap with $V_{\epsilon}$ as the ``reference code''. This term is a ``good'' encoding map  because it best corrects any single qubit erasure errors exactly. It is also a natural choice for the reference code when $\alpha\approx \beta$, as it is guaranteed to have the largest overlap with $V_{\epsilon}$.  Intuitively, we can think of this reference code as the QFT-on-curved-background approximation of a gravitating theory when we ignore the subleading effects of $G_N$. For the sake of concreteness, we have chosen each $V_i$ in Figure~\ref{fig:4skewed} to be an encoding isometry of a stabilizer code. 

\subsubsection{Lessons from Holographic Toy Models}

Although we do not yet have an explicit AQECC example with emergent flat geometry, one can distill a number of useful lessons from a known AQECC with emergent hyperbolic geometry, which also serves as a toy model for AdS/CFT. 

To construct such a model, let us replace the exact erasure correction codes (the yellow 4-qubit code tensors) in Figure~\ref{fig:holography}b by the approximate codes in Figure~\ref{fig:4skewed}, then one obtains a new encoding map (Figure~\ref{fig:holography}c) for an approximate quantum error correction code. 
One can verify  that the entropic relation now indeed becomes state-dependent where this dependence is controlled by the size of $\epsilon$ \cite{ABSC}. Generically, 
\begin{equation}
    |\tilde{\psi}\rangle \rightarrow |\tilde{\sigma}\rangle \implies F(\delta S_{geom})=\delta S_{matter} \ne 0 
\end{equation} 
for some function $F$.

In addition to getting us closer to our needed entropic relation (\ref{eqn:hamconstr}), the approximate code construction also seems to give rise to other features of gravity. 
One remarkable observation is that the logical degrees of freedoms that appear to be living in independent tensor factors are not truly independent --- instead, bulk operators acting on one region can impact another region that is spacelike separated\cite{Donnelly:2016rvo}. In particular, one can conclude that for two spacelike separated sites $x_1\ne x_2$ in the emergent tensor network geometry, the physical representation of logical operators $\tilde{Q}_{x_1}, \tilde{Q}'_{x_2}$ do not generally commute, but can have non-commutativity controlled by the ``noise'' parameter $\epsilon$, i.e., $[\tilde{Q}_{x_1}, \tilde{Q}'_{x_2}]\sim G(\epsilon)$. Here we take $G$ to be some function that satisfies $G(x\rightarrow 0)\rightarrow 0$.

The quantum information theoretic reason behind this is simple --- although the bulk or logical qubits appear as independent degrees of freedom, as shown by the red dots in Figure~\ref{fig:holography}, they actually overlap \cite{overlap} with each other by an amount related to the noise parameter $\epsilon$ \cite{Cao:2020uvb}. We can also understand it from the perspective that the mapping ~(\ref{eqn:nonisoencode})  is generically non-isometric\footnote{Superpositions of isometries need not be isometric.}. Therefore, orthogonal states in the logical subspace are not mapped to orthogonal states in the physical Hilbert space, i.e., 

\begin{equation}
    \langle \tilde{i}|\tilde{j}\rangle= \langle i |V^{\dagger}_{\epsilon}V_{\epsilon}|j\rangle \propto \langle i|(I+O(\epsilon)) |j\rangle = \delta_{ij}+O_{ij}(\epsilon), 
\end{equation}

where $O(\epsilon)$ denotes the correction terms that depend on the detailed structures of $V_{\epsilon}$ and vanishes as $\epsilon \rightarrow 0$. Similarly, even though $[Q_{x_1},Q_{x_2}]=0$, their commutator under the non-isometric mapping $[\tilde{Q}_{x_1},\tilde{Q}_{x_2}] = [V_{\epsilon}Q_{x_1}V_{\epsilon}^{\dagger},V_{\epsilon}Q_{x_2}V_{\epsilon}^{\dagger}]$ need not vanish because of the extra terms involving $O(\epsilon)$. Qualitatively, this is indeed what we expect to see in gravity as well. It is well-known that the Hilbert space for perturbative quantum gravity is not factorizable\cite{Donnelly:2015hta, Donnelly:2016rvo}. For matter coupled to gravitons, the gauge invariant dressed operators have non-vanishing commutators proportional to $G_N$. Here with the AQECC, we are finding a similar situation with $G_N\leftrightarrow \epsilon$, where these tiny overlaps are created by skewing the codes.

\begin{figure}
    \centering
    \includegraphics[width=0.65\linewidth]{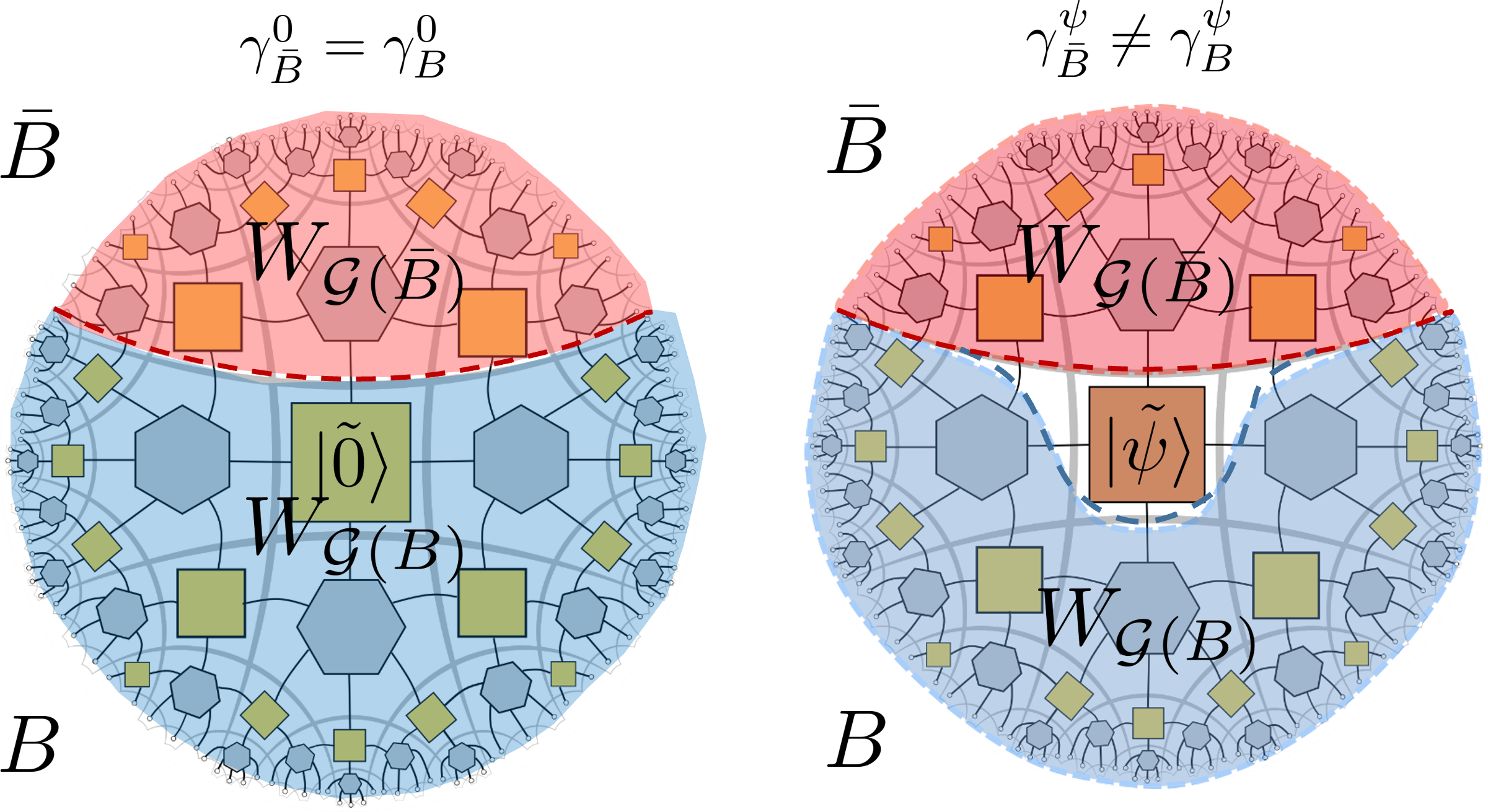}
    \caption{Changes in the bulk state induces changes in the emergent geometry as defined by  the entanglement wedge and the Greedy geodesic. Geodesics $\gamma_B,\gamma_{\bar{B}}$ are anchored on the same two boundary points. }
    \label{fig:backreaction}
\end{figure}

The following graphically intuitive observation is more specific to the particular tensor network model constructed for AdS/CFT. Nevertheless, we include it in our discussion to provide a more geometric perspective.  In the holographic model (Figure~\ref{fig:holography}c) introduced in \cite{ABSC}, one can also see the apparent similarity between gravitational backreaction and dependence on the logical state. For example, consider the ``vacuum'' all zero logical state represented by the yellow squares in Figure~\ref{fig:backreaction} (left). The emergent geodesics $\gamma_B^0, \gamma_{\bar{B}}^0$ anchored on two points on the boundary can be found through a Greedy decoding process. They correspond to the dash line that mark the boundary of the entanglement wedges coloured in blue and pink. In this case, the two geodesics from the two wedges coincide, which is consistent with our expectation when the background geometry is pure AdS (hyperbolic). 

However, when the state at the center is changed to a superposition $|\tilde{\psi}\rangle = a|\tilde{0}\rangle+b|\tilde{1}\rangle$, then the same decoding procedure produces different wedges that are now demarcated by two non-coincidental geodesics, skirting around the central region. This is similar to having a massive object inserted at the center of AdS\footnote{Note the difference between this exercise and general expectations in AdS/CFT. Here the computational basis states $|\tilde{0}\rangle, |\tilde{1}\rangle$ encoding the classical bit of information is playing the role of pure states in holography, as the recovery is complementary. However, a state in a superposition is playing the role of a mixed state because any subsystem can only access the Z code subalgebra. This reduces $|\tilde{\psi}\rangle$ to a mixed state when expressed as a state in that subalgebra.} (Figure \ref{fig:backreaction} right). 

Finally, we also see that the noise parameter $\epsilon$ is somewhat analogous to $G_N$, which controls the emergence of these ``gravitational effects''. Similar to recovering EFT on curved background by taking $G_N\rightarrow 0$, we recover an exact stabilizer QECC model (Figure~\ref{fig:holography}b) in the $\epsilon\rightarrow 0$ limit, where the ``emergent gravitational effects'' also vanish. 
In addition, if $\epsilon$ is large, supposedly analogous to the scenario in which $G_N$ is large and gravity is highly quantum, then we find that we can not identify a dominant ``reference code'' like $V_0$ in (\ref{eqn:nonisoencode}). Since the reference code can be thought of as a particular background (network) geometry, here we have a highly quantum geometry that is a macroscopic superposition of potentially different network geometries each defined by a different encoding map. In this case, there is no obvious way for us to recover a ``semi-classical'' picture as there is no unique choice of a dominant reference code $V_0$.

\subsection{Generalizations and Ways forward}
In summary, in order to also produce the spatial part of linearized Einstein's equations around flat space, one needs the additional constraint,

\begin{enumerate}[resume]
    \item (Modified Entanglement Equilibrium Condition) 
    \begin{equation}
        \delta S_{\rm geom}(\chi_{B_2}) +\delta S_{\rm matter}(\rho_{B_1}) =0
        \label{eqn:meec}
    \end{equation}
    for certain bipartitions of the Hilbert space factors into $B$ and $\bar{B}$, which is described in detail in \cite{BEG}. 
\end{enumerate}

However, this constraint is not satisfied by any existing exact erasure correction codes. There is also suspicion whether it could be satisfied at all by any stabilizer codes, which are popular for building practical QECCs\cite{Pollack:2021yij}. However, we showed that by making these nice codes a little worse through coherent noise injection, we obtain approximate QECCs that are non-stabilizer codes which can satisfy condition $F(\delta S_{\rm geom})=\delta S_{\rm matter}$.

Therefore, to build up a better model with emergent near-flat geometry that satisfies the linearized Hamiltonian constraint, it remains to (i) identify codes for which $F(\delta S_{\rm geom})=-\delta S_{\rm geom}$ and (ii) construct explicit examples that have  emergent flat geometry, as opposed to the hyperbolic one we have seen so far\footnote{Considering that Einstein gravity may be modified to incorporate higher order terms, it is also interesting identify a code for which $F$ has a Taylor expansion whose leading linear contribution is $-\delta S_{\rm geom}$.}. Fortunately, because the QECC themselves are not tied to any pre-existing notion of background geometries, there is no obvious obstacle for generalizations beyond emergent AdS geometries. A possible candidate for such an exact QECC can be constructed using 5-qubit codes in \cite{Cao:2021ibt}. Then by skewing the 5-qubit code tensors, one can attempt a construction for AQECCs with near-flat geometry. Alternatively, one can directly construct an AQECC with near-flat emergent geometry using random tensor networks \cite{Hayden:2016cfa} in the large bond dimension limit. 





\section{Discussion}
\label{sec:discussion}
\subsection{Towards Emergent Einstein Gravity in Minkowski Spacetime}
Once we can reproduce the linearized Hamiltonian constraint following the steps above, we then need the following additional ingredients to recover the full linearized Einstein's equations on Minkowski background.

\begin{enumerate}[resume]
    \item There exists a consistent dynamical theory, e.g. a Hamiltonian or quantum circuit, that can generate a sequence of states each admitting an emergent spatial geometry, which can be organized to create a spacetime geometry.
    \item The overall theory is Lorentz invariant in the appropriate limit and that the above conditions hold for all constant time slices in the emergent background spacetime.
\end{enumerate}


Let us briefly discuss how the last two conditions can be met. One possible avenue to incorporate dynamics involves picking a Hamiltonian. Some quantum codes such as stabilizer codes admit natural constructions of stabilizer Hamiltonians for which states in the code subspace live in the ground space. There are also numerous quantum manybody models whose low energy subspace naturally correspond to approximate QECCs\cite{SY,kitaev,Brandao2019}. Alternatively, time may be emergent. Some proposals have suggested the modular flow as a possible substitute for proper time\cite{tth}. It has also been made more concrete recently in the context of AdS/CFT\cite{Lampros}. Other possibilities may involve tensor networks or  quantum circuits generated based on some background spacetime\cite{Qi:2018shh,dSMERA,Osborne,Milsted:2018san}.

General understanding of how Lorentz symmetries may emerge is still lacking. Traditionally, there are known obstacles in emerging Lorentz symmetries from the kind of systems we consider as there is no finite dimensional unitary irreducible representation of the Lorentz group. Although one might have hoped that Lorentz symmetry is only approximate and breaks down in some high energy regime, the lack of observational signature also renders this line of thinking difficult. However, it may be possible to identify a class or sequence of finite groups of increasing order that approximates the Lorentz group with increasingly better accuracy in the large size limit. A potentially relevant work has been explored by \cite{Hu:2017rsp} in the context of recovering conformal symmetries using finite dimensional systems. 

Finally, we need to move beyond linearized gravity. A few works by \cite{Jacobson:2015hqa,Faulkner:2017tkh} are able to attain the non-linear Einstein's equations with different techniques and assumptions, which are worth further exploration.
\subsection{Summary}
In this note, we reviewed the recent gravitize QM proposal in light of using QECCs as concrete implementations of emergence maps. By identifying the right class of (approximate) QECCs, we hope to derive from first principles not only emergent geometry beyond AdS but also gravity from quantum information constraints. We showed that many desirable properties needed for the gravitize QM proposal can be naturally found in QECCs that are sufficiently generic. Combined with techniques introduced in \cite{Cao2017,BEG}, we provided a cursory guide that takes us from quantum states in Hilbert spaces to emergent matter field on fixed background geometries. 

However, in order for gravitational effects to emerge, we further restrict ourselves to approximate QECCs. Interestingly, such systems can be easily constructed when we subject our existing QECC models to coherent noise. These codes are the opposite of what one typically wants for fault-tolerant quantum computing, but are more natural to implement on noisy quantum devices such as the ones being constructed in the near term\cite{NISQ}. By studying an AQECC model developed for holography, we learned that the level of noise is analogous to the strength of the gravitational coupling. Furthermore, the injected noise introduces the necessary entropic relations for gravitational backreactions and renders bulk operators weakly non-local, consistent with our intuitions from quantum gravity. We hope to generalize these lessons to a class of codes that have emergent near-flat geometry and satisfy the modified entanglement equilibrium condition~(\ref{eqn:meec}). 

Notably, quantum noise here plays an essential role for introducing some desirable features of gravity. Instead of treating noise as a bug that we wish to remove in quantum computing, we may use the quantum noise in near-term quantum devices to our advantage. One possibility would be to implement such kind of ``noisy'' quantum codes with emergent geometries on devices dominated by coherent noise and simulate certain aspects of quantum gravity. Further work is needed to establish a more robust experimental direction for studying quantum gravity in realistic quantum devices.

\section*{Acknowledgement}
C.C. would like to thank the organizers of \textit{The Quantum and The Gravity 2021 Workshop} for which this note is prepared as a part of the proceeding. C.C. acknowledges the support by the U.S. Department of Defense and NIST through the Hartree Postdoctoral Fellowship at QuICS, by the Simons Foundation as part of the It From Qubit Collaboration, and by the DOE Office of Science, Office of High Energy Physics, through the grant DE-SC0019380. 

\bibliographystyle{unsrt} 
\bibliography{ref} 

\end{document}